# A Further Generalized Lagrangian Density and Its Special Cases

**Fang-Pei Chen**

Department of Physics, Dalian University of Technology, Dalian 116024, China.

E-mail: chenfap@dlut.edu.cn

**Abstract** By summarizing and extending the Lagrangian densities of the general relativity and the Kibble's gauge theory of gravitation, a further generalized Lagrangian density for a gravitational system is obtained and analyzed in greater detail, which can be used for studying more extensive range of gravitation. Many special cases can be derived from this generalized Lagrangian density, their general characters and peculiarities will be briefly described.

## 1. Introduction

In the theory of special relativity, the Lagrangian of matter field $\psi(x)$ can be denoted by the functional form:

$$L_M(x) = L_M[\psi(x); \psi_{,\mu}(x)] \qquad (1)$$

where $\psi_{,\mu}(x)$ is the ordinary derivative of $\psi(x)$. It is well known that, in the relativistic theories of gravitation, the Lagrangian density of matter field must be denoted by the functional form [1,2]:

$$\sqrt{-g}\, L_M(x) = \sqrt{-g}\, L_M[\psi(x); \psi_{|\mu}(x); h^i_{\cdot\mu}(x)] \qquad (2)$$

where $\psi_{|\mu}(x)$ is the covariant derivative of $\psi(x)$:

$$\psi_{|\mu}(x) = \psi_{,\mu}(x) + \frac{1}{2}\Gamma^{ij}_{\cdot\cdot\mu}(x)\sigma_{ij}\psi(x) \qquad (3)$$

For the Kibble's gauge theory of gravitation [1], the frame connection $\Gamma^{ij}_{\cdot\cdot\mu}(x)$ is independent field variables and the torsion must appear in the space-time. In this case the Eq.(2) can be generalized as

$$\sqrt{-g(x)}\, L_M(x) = \sqrt{-g(x)}\, L_M[\psi(x); \psi_{,\lambda}(x); h^i_{\cdot\mu}(x); \Gamma^{ij}_{\cdot\cdot\mu}(x)] \qquad (4)$$

For the relativistic theories of gravitation in the space-time without torsion, $\Gamma^{ij}_{..\mu}(x)$ should not be independent field variables; and it will be proven in the Appendix that

$$\Gamma^{ij}_{..\mu} = \frac{1}{2}\eta^{jk} h^{\nu}_k (h^i_{.\mu,\nu} - h^i_{.\nu,\mu}) + \frac{1}{2}\eta^{id} h^{\sigma}_d (h^j_{.\sigma,\mu} - h^j_{.\mu,\sigma}) \\ + \frac{1}{2}\eta^{jk} h^{\nu}_k \eta^{id} h^{\sigma}_d \eta_{ab} h^b_{.\mu} (h^a_{.\sigma,\nu} - h^a_{.\nu,\sigma}) \quad (5)$$

In this case the Eq.(2) can be generalized as

$$\sqrt{-g(x)}\, L_M(x) = \sqrt{-g(x)}\, L_M[\psi(x); \psi_{,\lambda}(x); h^i_{.\mu}(x); h^i_{.\mu,\nu}(x)] \quad (6)$$

Since the great majority of the fundamental matter fields are spinors, it is necessary to use tetrad field $h^i_{.\mu}(x)$ [2]. The metric field $g_{\mu\nu}(x)$ is expressed as $g_{\mu\nu}(x) = h^i_{.\mu}(x) h^j_{.\nu}(x) \eta_{ij}$, from which we have

$$h^{\mu}_i(x) = \eta_{ij}\, g^{\mu\nu}(x) h^j_{.\nu}(x)\ ; \quad h_{i\nu,\lambda}(x) = \frac{\partial}{\partial x^\lambda} h_{i\nu}(x)\ ; \quad \text{etc.}$$

In the relativistic theories of gravitation,

$$\sqrt{-g}\, L_G(x) = \frac{\sqrt{-g}}{16\pi G} R(x) \quad (7)$$

is always adopted as the Lagrangian density of gravitational field [1,2]. For the Kibble's gauge theory of gravitation [1], Eq.(7) can be generalized as

$$\sqrt{-g(x)}\, L_G(x) = \sqrt{-g(x)}\, L_G[h^i_{.\mu}(x); \Gamma^{ij}_{..\mu}(x); \Gamma^{ij}_{..\mu,\lambda}(x)] \quad (8)$$

For the relativistic theories of gravitation in the space-time without torsion (e.g. general relativity), after using Eq.(5), Eq.(7) can be generalized as [9]

$$\sqrt{-g(x)}\, L_G(x) = \sqrt{-g(x)}\, L_G[h^i_{.\mu}(x); h^i_{.\mu,\lambda}(x); h^i_{.\mu,\lambda\sigma}(x)] \quad (9)$$

In this paper, in order to conduct an indepth study on the general character and the peculiarity of Lagrangian densities for some relativistic theories of gravitation, Eqs.(4,6) will be extended to the following expression:

$$\sqrt{-g(x)}\, L_M(x) = \sqrt{-g(x)}\, L_M[\psi(x); \psi_{,\lambda}(x); h^i_{.\mu}(x); h^i_{.\mu,\lambda}(x); \Gamma^{ij}_{..\mu}(x); \Gamma^{ij}_{..\mu,\lambda}(x);] \quad (10)$$

and Eqs.(8,9) will be extended to the following expression:

$$\sqrt{-g(x)}\, L_G(x) = \sqrt{-g(x)}\, L_G[h^i_{.\mu}(x); h^i_{.\mu,\lambda}(x); \Gamma^{ij}_{..\mu}(x); \Gamma^{ij}_{..\mu,\lambda}(x)] \quad (11)$$

We will name $\sqrt{-g(x)} L(x) = \sqrt{-g(x)} L_M(x) + \sqrt{-g(x)} L_G(x)$ （where $\sqrt{-g(x)} L_M(x)$ and $\sqrt{-(g)} L_G(x)$ are denoted by Eq.(10) and Eq.(11) respectively） as a further generalized Lagrangian density. $\psi(x)$ represents the matter field, $h^i_{.\mu}(x)$ and $\Gamma^{ij}_{..\mu}(x)$ represent the gravitational fields. This further generalized Lagrangian density is significantly more general than the Lagrangian densities denoted by Eqs.(4,6) and Eqs.(8,9)

It must be indicated that, apart from describing a gravitational system with torsion, this further generalized Lagrangian density (*i.e.* Eqs.(10,11) ) can be used also to describe a gravitational system without torsion. If Eqs.(10,11) are used to describe a gravitational system without torsion, it must be noted that $\Gamma^{ij}_{..\mu}(x)$ is function of $h^i_{.\mu}(x), h^i_{.\mu,\lambda}(x)$, and $\Gamma^{ij}_{..\mu,\lambda}(x)$ is function of $h^i_{.\mu}(x), h^i_{.\mu,\lambda}(x), h^i_{.\mu,\lambda\sigma}(x)$. So the Eq.(10) can be expressed as

$$\sqrt{-g(x)} L_M(x) = \sqrt{-g(x)} L_M [\psi(x); \psi_{,\lambda}(x); h^i_{.\mu}(x); h^i_{.\mu,\lambda}(x);$$
$$\Gamma^{ab}_{..\alpha}[h^i_{.\mu}(x); h^i_{.\mu,\lambda}(x)]; \Gamma^{ab}_{..\alpha,\beta}[h^i_{.\mu}(x); h^i_{.\mu,\lambda}(x); h^i_{.\mu,\lambda\sigma}(x)]] \quad (12)$$
$$= \sqrt{-g(x)} L^*_M [\psi(x); \psi_{,\lambda}(x); h^i_{.\mu}(x); h^i_{.\mu,\lambda}(x); h^i_{.\mu,\lambda\sigma}]$$

and the Eq.(11) can be expressed as

$$\sqrt{-g(x)} L_G(x) = \sqrt{-g(x)} L_G [h^i_{.\mu}(x); h^i_{.\mu,\lambda}(x);$$
$$\Gamma^{ab}_{..\alpha}[h^i_{.\mu}(x); h^i_{.\mu,\lambda}(x)]; \Gamma^{ab}_{..\alpha,\beta}[h^i_{.\mu}(x); h^i_{.\mu,\lambda}(x); h^i_{.\mu,\lambda\sigma}(x)]] \quad (13)$$
$$= \sqrt{-g(x)} L^*_G [h^i_{.\mu}(x); h^i_{.\mu,\lambda}(x); h^i_{.\mu,\lambda\sigma}(x)]$$

Evidently Eq.(6) is a special case of Eq.(12) when $h^i_{.\mu,\lambda\sigma}(x) \equiv 0$, Eq.(9) is equivalent to Eq.(13).

For the relativistic theories of gravitation in the space-time with torsion, besides Eq.(4), the following Lagrangian densities

$$\sqrt{-g(x)} L_M(x) = \sqrt{-g(x)} L_M [\psi(x); \psi_{,\lambda}(x); h^i_{.\mu}(x); h^i_{.\mu,\lambda}(x); \Gamma^{ij}_{..\mu}(x)] \quad (14)$$

$$\sqrt{-g(x)} L_M(x) = \sqrt{-g(x)} L_M [\psi(x); \psi_{,\lambda}(x); h^i_{.\mu}(x); \Gamma^{ij}_{..\mu}(x); \Gamma^{ij}_{..\mu,\lambda}(x)] \quad (15)$$

are also the special cases of Eq.(10). By means of studying the further generalized Lagrangian density and its special cases, their general character and peculiarity can be shown clearly.

The further generalized Lagrangian density summarizes many properties of various theories of gravitation. Below we shall prove that, Eq.(10) and Eq.(11) can be rewritten as

$$\sqrt{-g(x)}\, L_M(x) = \sqrt{-g(x)}\, L_M^\#(x) = \sqrt{-g(x)}\, L_M^\#[\psi(x); \psi_{|\mu}(x); R^{ij}_{..\mu\nu}(x)\, ; T^i_{.\mu\nu}(x)\, ; h^i_{.\mu}(x)\, ] \qquad (16)$$

and

$$\sqrt{-g(x)}\, L_G(x) = \sqrt{-g(x)}\, L_G^\#(x) = \sqrt{-g(x)}\, L_G^\#[R^{ij}_{..\mu\nu}(x)\, ; T^i_{.\mu\nu}(x)\, ; h^i_{.\mu}(x)\, ] \qquad (17)$$

$R^{ij}_{..\mu\nu}$ is the curvature tensor with mixed indexs,

$$R^{ij}_{..\mu\nu} = \Gamma^{ij}_{..\nu,\mu} - \Gamma^{ij}_{..\mu,\nu} + \delta^i_m \eta_{nk}(\Gamma^{mn}_{..\mu}\Gamma^{kj}_{..\nu} - \Gamma^{mn}_{..\nu}\Gamma^{kj}_{..\mu}); \qquad (18)$$

$T^i_{.\mu\nu}$ is the torsion tensor with mixed indexes,

$$T^i_{.\mu\nu} = \frac{1}{2}\{h^i_{.\mu,\nu} - h^i_{.\nu,\mu} + \delta^i_m \eta_{nk}(\Gamma^{mn}_{..\nu} h^k_\mu - \Gamma^{mn}_{..\mu} h^k_\nu)\} \qquad (19)$$

The physical meaning of Eq.(16) is that the gravitational fields could act on the matter field only through covariant derivative, curvature of space-time, and torsion of space-time. Therefore the forms of couplings between the gravitational fields and matter field might be $\Gamma^{ij}_{..\mu}\sigma_{ij}\psi$, $R|\psi|^2$, $R^{ij}_{..\mu\nu} R_{ij}^{..\mu\nu}|\psi|^2$ or $T^i_{.\mu\nu} T_i^{.\mu\nu}|\psi|^2$ etc. The coupling $\Gamma^{ij}_{..\mu}\sigma_{ij}\psi$ contained in the covariant derivative $\psi_{|\mu}(x)$ is called the minimal coupling, which is well known in the general relativity and the gauge theory of gravitation. Eq.(16) tells us that in addition to the minimal coupling, there might be other complicated couplings in theory. The physical meaning of Eq.(17) is that the Lagrangian of gravitational field is composed of curvature tensor field and torsion tensor field. Because $L_G^\#(x)$ is both a coordinate scalar and a frame scalar, the possible terms involved in $L_G^{\#}(x)$ are scalars constructed from $R^{ij}_{..\mu\nu}(x), T^{i}_{.\mu\nu}(x), h^{i}_{.\mu}(x)$. Hence, the study of further generalized Lagrangian densities significantly extend the range of the studies of gravitation. If Eq.(17) is used to describe a gravitational system without torsion, then $T^{i}_{.\mu\nu}(x)=0$ and $R^{\sigma(\Gamma)}_{.\lambda\mu\nu} = R^{\sigma(\{\})}_{.\lambda\mu\nu}$; the possible terms involved in $L_G(x)$ are only the scalar curvature $\overset{\{\}}{R} = h^{.\mu}_i h^{.j}_j R^{ij}_{..\mu\nu}$ and its power such as $\overset{\{\}}{R}{}^2$ …. Considering other requirements [2], $L_G(x) = \frac{1}{16\pi G}[R(x) + 2\lambda]$ is chosen in general relativity.

We shall discuss the above problems in the following sections.

## 2. The symmetry of the Lagrangian densities for a gravitational system

Symmetries exist universally in physical systems. We suppose that one fundamental symmetry of a gravitational system is that the action integrals

$$I_M = \int \sqrt{-g(x)} L_M(x) d^4x \qquad I_G = \int \sqrt{-g(x)} L_G(x) d^4x$$

and

$$I = I_M + I_G = \int \sqrt{-g(x)} (L_M(x) + L_G(x)) d^4x$$

satisfy $\delta I_M = 0$, $\delta I_G = 0$ and $\delta I = 0$ respectively under the following two simultaneous transformations [1,4]:

(1), the infinitesimal general coordinate transformation

$$x^\mu \to x'^\mu = x^\mu + \xi^\mu(x) \tag{20}$$

(2), the local Lorentz transformation of tetrad frame

$$e_i(x) \to e'_i(x') = e_i(x) - \varepsilon^{mn}(x) \delta^j_m \eta_{ni} e_j(x) \tag{21}$$

The sufficient condition of an action integral $I = \int \sqrt{-g(x)} L(x) d^4x$ being $\delta I = 0$ under above transformations is [1,10]:

$$\delta_0(\sqrt{-g}L) + (\xi^\mu \sqrt{-g}L)_{,\mu} \equiv 0 \tag{22}$$

where $\delta_0$ represents the variation at a fixed value of $x$. For the most generalized Lagrangian density we have

$$\delta_0(\sqrt{-g}L_M) = \frac{\partial(\sqrt{-g}L_M)}{\partial \psi}\delta_0\psi + \frac{\partial(\sqrt{-g}L_M)}{\partial \psi_{,\lambda}}\delta_0\psi_{,\lambda} + \frac{\partial(\sqrt{-g}L_M)}{\partial h^i_{.\mu}}\delta_0 h^i_{.\mu}$$
$$+ \frac{\partial(\sqrt{-g}L_M)}{\partial h^i_{.\mu,\lambda}}\delta_0 h^i_{.\mu,\lambda} + \frac{\partial(\sqrt{-g}L_M)}{\partial \Gamma^{ij}_{..\mu}}\delta_0\Gamma^{ij}_{..\mu} + \frac{\partial(\sqrt{-g}L_M)}{\partial \Gamma^{ij}_{..\mu,\lambda}}\delta_0\Gamma^{ij}_{..\mu,\lambda} \tag{23}$$

$$\delta_0(\sqrt{-g}L_G) = \frac{\partial(\sqrt{-g}L_G)}{\partial h^i_{.\mu}}\delta_0 h^i_{.\mu} + \frac{\partial(\sqrt{-g}L_G)}{\partial h^i_{.\mu,\lambda}}\delta_0 h^i_{.\mu,\lambda}$$
$$+ \frac{\partial(\sqrt{-g}L_G)}{\partial \Gamma^{ij}_{..\mu}}\delta_0\Gamma^{ij}_{..\mu} + \frac{\partial(\sqrt{-g}L_G)}{\partial \Gamma^{ij}_{..\mu,\lambda}}\delta_0\Gamma^{ij}_{..\mu,\lambda} \tag{24}$$

Let $\Lambda = L_M$ or $\Lambda = L_G$ or $\Lambda = L_M + L_G$, because of the independent arbitrariness of $\varepsilon^{mn}_{,}(x)$, $\varepsilon^{mn}_{,\lambda}(x)$, $\varepsilon^{mn}_{,\lambda\sigma}(x)$, $\xi^\alpha(x)$, $\xi^\alpha_{,\mu}(x)$ and $\xi^\alpha_{,\mu\lambda}(x)$, It is not difficult to derive the following identities [3] ( if $\Lambda = L_G$, $\frac{\partial(\sqrt{-g}\Lambda)}{\partial \psi} = 0$, $\frac{\partial(\sqrt{-g}\Lambda)}{\partial \psi_{,\lambda}} = 0$ ):

$$\frac{1}{2}\frac{\partial(\sqrt{-g}\Lambda)}{\partial \psi}\sigma_{mn}\psi + \frac{1}{2}\frac{\partial(\sqrt{-g}\Lambda)}{\partial \psi_{,\lambda}}\sigma_{mn}\psi_{,\lambda} + \frac{\partial(\sqrt{-g}\Lambda)}{\partial h^m_{.\mu}}h_{n\mu}$$
$$+ \frac{\partial(\sqrt{-g}\Lambda)}{\partial h^m_{.\mu,\lambda}}h_{n\mu,\lambda} + 2\frac{\partial(\sqrt{-g}\Lambda)}{\partial \Gamma^{km}_{..\mu}}\Gamma^k_{.n\mu} + 2\frac{\partial(\sqrt{-g}\Lambda)}{\partial \Gamma^{km}_{..\mu,\lambda}}\Gamma^k_{.n\mu,\lambda} = 0 \tag{25}$$

$$\frac{1}{2}\frac{\partial(\sqrt{-g}\Lambda)}{\partial \psi_{,\lambda}}\sigma_{mn}\psi + \frac{\partial(\sqrt{-g}\Lambda)}{\partial h^{m}_{.\mu,\lambda}}h_{n\mu} + 2\frac{\partial(\sqrt{-g}\Lambda)}{\partial \Gamma^{km}_{..\mu,\lambda}}\Gamma^{k}_{.n\mu} = \frac{\partial(\sqrt{-g}\Lambda)}{\partial \Gamma^{mn}_{..\lambda}} \qquad (26)$$

$$\frac{\partial(\sqrt{-g}\Lambda)}{\partial \Gamma^{mn}_{..\mu,\nu}} = -\frac{\partial(\sqrt{-g}\Lambda)}{\partial \Gamma^{mn}_{..\nu,\mu}} \qquad (27)$$

$$\frac{\partial(\sqrt{-g}\Lambda)}{\partial \psi}\psi_{,\alpha} + \frac{\partial(\sqrt{-g}\Lambda)}{\partial \psi_{,\lambda}}\psi_{,\lambda\alpha} + \frac{\partial(\sqrt{-g}\Lambda)}{\partial h^{i}_{.\mu}}h^{i}_{.\mu,\alpha}$$
$$+ \frac{\partial(\sqrt{-g}\Lambda)}{\partial h^{i}_{.\mu,\lambda}}h^{i}_{.\mu,\alpha\lambda} + \frac{\partial(\sqrt{-g}\Lambda)}{\partial \Gamma^{ij}_{..\mu}}\Gamma^{ij}_{..\mu,\alpha} + \frac{\partial(\sqrt{-g}\Lambda)}{\partial \Gamma^{ij}_{..\mu,\lambda}}\Gamma^{ij}_{..\mu,\alpha\lambda} - (\sqrt{-g}\Lambda)_{,\alpha} = 0 \qquad (28)$$

$$\frac{\partial(\sqrt{-g}\Lambda)}{\partial \psi_{,\lambda}}\psi_{,\alpha} + \frac{\partial(\sqrt{-g}\Lambda)}{\partial h^{i}_{.\lambda}}h^{i}_{.\alpha} + \frac{\partial(\sqrt{-g}\Lambda)}{\partial h^{i}_{.\mu,\lambda}}(h^{i}_{.\mu,\alpha} - h^{i}_{.\alpha,\mu})$$
$$+ \frac{\partial(\sqrt{-g}\Lambda)}{\partial \Gamma^{ij}_{..\lambda}}\Gamma^{ij}_{..\alpha} + \frac{\partial(\sqrt{-g}\Lambda)}{\partial \Gamma^{ij}_{..\mu,\lambda}}(\Gamma^{ij}_{..\mu,\alpha} - \Gamma^{ij}_{..\alpha,\mu}) - \sqrt{-g}\Lambda \delta^{\lambda}_{\alpha} = 0 \qquad (29)$$

$$\frac{\partial(\sqrt{-g}\Lambda)}{\partial h^{i}_{.\mu,\lambda}}h^{i}_{.\alpha} + \frac{\partial(\sqrt{-g}\Lambda)}{\partial \Gamma^{ij}_{..\mu,\lambda}}\Gamma^{ij}_{..\alpha} = 0 \qquad (30)$$

From Eq.(30) and Eq.(27), it is found that there must exist another identity:

$$\frac{\partial(\sqrt{-g}\Lambda)}{\partial h^{i}_{.\mu,\nu}} = -\frac{\partial(\sqrt{-g}\Lambda)}{\partial h^{i}_{.\nu,\mu}} \qquad (31)$$

It will be shown bellow that many properties of a gravitational system can be derived from the above identities.

When Eqs.(10,11) are used to describe a gravitational system without torsion, from Eq.(12) we have

$$\delta_0(\sqrt{-g}\,L_M) = \frac{\partial(\sqrt{-g}\,L_M)}{\partial \psi}\delta_0\psi + \frac{\partial(\sqrt{-g}\,L_M)}{\partial \psi_{,\lambda}}\delta_0\psi_{,\lambda} + \left(\frac{\partial(\sqrt{-g}\,L_M)}{\partial h^i_{.\mu}}\right)_\Gamma \delta_0 h^i_{.\mu}$$

$$+\left(\frac{\partial(\sqrt{-g}\,L_M)}{\partial h^i_{.\mu,\lambda}}\right)_\Gamma \delta_0 h^i_{.\mu,\lambda} + \frac{\partial(\sqrt{-g}\,L_M)}{\partial \Gamma^{ab}_{..\alpha}}\frac{\partial \Gamma^{ab}_{..\alpha}}{\partial h^i_{.\mu}}\delta_0 h^i_{.\mu} + \frac{\partial(\sqrt{-g}\,L_M)}{\partial \Gamma^{ab}_{..\alpha}}\frac{\partial \Gamma^{ab}_{..\alpha}}{\partial h^i_{.\mu,\lambda}}\delta_0 h^i_{.\mu,\lambda}$$

$$+\frac{\partial(\sqrt{-g}\,L_M)}{\partial \Gamma^{ab}_{..\alpha,\beta}}\frac{\partial \Gamma^{ab}_{..\alpha,\beta}}{\partial h^i_{.\mu}}\delta_0 h^i_{.\mu} + \frac{\partial(\sqrt{-g}\,L_M)}{\partial \Gamma^{ab}_{..\alpha,\beta}}\frac{\partial \Gamma^{ab}_{..\alpha,\beta}}{\partial h^i_{.\mu,\lambda}}\delta_0 h^i_{.\mu,\lambda} + \frac{\partial(\sqrt{-g}\,L_M)}{\partial \Gamma^{ab}_{..\alpha,\beta}}\frac{\partial \Gamma^{ab}_{..\alpha,\beta}}{\partial h^i_{.\mu,\lambda\sigma}}\delta_0 h^i_{.\mu,\lambda\sigma}$$

$$=\delta_0(\sqrt{-g}\,L^*_M) = \frac{\partial(\sqrt{-g}\,L^*_M)}{\partial \psi}\delta_0\psi + \frac{\partial(\sqrt{-g}\,L^*_M)}{\partial \psi_{,\lambda}}\delta_0\psi_{,\lambda} + \frac{\partial(\sqrt{-g}\,L^*_M)}{\partial h^i_{.\mu}}\delta_0 h^i_{.\mu}$$

$$+\frac{\partial(\sqrt{-g}\,L^*_M)}{\partial h^i_{.\mu,\lambda}}\delta_0 h^i_{.\mu,\lambda} + \frac{\partial(\sqrt{-g}\,L^*_M)}{\partial h^i_{.\mu,\lambda\sigma}}\delta_0 h^i_{.\mu,\lambda\sigma}$$

(32)

where $\left(\frac{\partial}{\partial}\right)_\Gamma$ denote the partial derivative at the constant values of $\Gamma^{ab}_{..\alpha}(x)$ and $\Gamma^{ab}_{..\alpha,\beta}(x)$. Hence we get

$$\frac{\partial(\sqrt{-g}\,L^*_M)}{\partial \psi} = \frac{\partial(\sqrt{-g}\,L_M)}{\partial \psi} \tag{33}$$

$$\frac{\partial(\sqrt{-g}\,L^*_M)}{\partial \psi_{,\lambda}} = \frac{\partial(\sqrt{-g}\,L_M)}{\partial \psi_{,\lambda}} \tag{34}$$

$$\frac{\partial(\sqrt{-g}\,L^*_M)}{\partial h^i_{.\mu}} = \left(\frac{\partial(\sqrt{-g}\,L_M)}{\partial h^i_{.\mu}}\right)_\Gamma + \frac{\partial(\sqrt{-g}\,L_M)}{\partial \Gamma^{ab}_{..\alpha}}\frac{\partial \Gamma^{ab}_{..\alpha}}{\partial h^i_{.\mu}} + \frac{\partial(\sqrt{-g}\,L_M)}{\partial \Gamma^{ab}_{..\alpha,\beta}}\frac{\partial \Gamma^{ab}_{..\alpha,\beta}}{\partial h^i_{.\mu}} \tag{35}$$

$$\frac{\partial(\sqrt{-g}\,L^*_M)}{\partial h^i_{.\mu,\lambda}} = \left(\frac{\partial(\sqrt{-g}\,L_M)}{\partial h^i_{.\mu,\lambda}}\right)_\Gamma + \frac{\partial(\sqrt{-g}\,L_M)}{\partial \Gamma^{ab}_{..\alpha}}\frac{\partial \Gamma^{ab}_{..\alpha}}{\partial h^i_{.\mu,\lambda}} + \frac{\partial(\sqrt{-g}\,L_M)}{\partial \Gamma^{ab}_{..\alpha,\beta}}\frac{\partial \Gamma^{ab}_{..\alpha,\beta}}{\partial h^i_{.\mu,\lambda}} \tag{36}$$

$$\frac{\partial(\sqrt{-g}\,L^*_M)}{\partial h^i_{.\mu,\lambda\sigma}} = \frac{\partial(\sqrt{-g}\,L_M)}{\partial \Gamma^{ab}_{..\alpha,\beta}}\frac{\partial \Gamma^{ab}_{..\alpha,\beta}}{\partial h^i_{.\mu,\lambda\sigma}} \tag{37}$$

From Eq.(13) we have

$$\delta_0(\sqrt{-g}\,L_G) = (\frac{\partial(\sqrt{-g}\,L_G)}{\partial h^i_{.\mu}})_\Gamma \delta_0 h^i_{.\mu} + (\frac{\partial(\sqrt{-g}\,L_G)}{\partial h^i_{.\mu,\lambda}})_\Gamma \delta_0 h^i_{.\mu,\lambda}$$

$$+ \frac{\partial(\sqrt{-g}\,L_G)}{\partial \Gamma^{ab}_{..\alpha}} \frac{\partial \Gamma^{ab}_{..\alpha}}{\partial h^i_{.\mu}} \delta_0 h^i_{.\mu} + \frac{\partial(\sqrt{-g}\,L_G)}{\partial \Gamma^{ab}_{..\alpha}} \frac{\partial \Gamma^{ab}_{..\alpha}}{\partial h^i_{.\mu,\lambda}} \delta_0 h^i_{.\mu,\lambda}$$

$$+ \frac{\partial(\sqrt{-g}\,L_G)}{\partial \Gamma^{ab}_{..\alpha,\beta}} \frac{\partial \Gamma^{ab}_{..\alpha,\beta}}{\partial h^i_{.\mu}} \delta_0 h^i_{.\mu} + \frac{\partial(\sqrt{-g}\,L_G)}{\partial \Gamma^{ab}_{..\alpha,\beta}} \frac{\partial \Gamma^{ab}_{..\alpha,\beta}}{\partial h^i_{.\mu,\lambda}} \delta_0 h^i_{.\mu,\lambda} + \frac{\partial(\sqrt{-g}\,L_G)}{\partial \Gamma^{ab}_{..\alpha,\beta}} \frac{\partial \Gamma^{ab}_{..\alpha,\beta}}{\partial h^i_{.\mu,\lambda\sigma}} \delta_0 h^i_{.\mu,\lambda\sigma}$$

$$= \delta_0(\sqrt{-g}\,L_G^*) = \frac{\partial(\sqrt{-g}\,L_G^*)}{\partial h^i_{.\mu}} \delta_0 h^i_{.\mu} + \frac{\partial(\sqrt{-g}\,L_G^*)}{\partial h^i_{.\mu,\lambda}} \delta_0 h^i_{.\mu,\lambda} + \frac{\partial(\sqrt{-g}\,L_G^*)}{\partial h^i_{.\mu,\lambda\sigma}} \delta_0 h^i_{.\mu,\lambda\sigma}$$

(38)

Hence we get

$$\frac{\partial(\sqrt{-g}\,L_G^*)}{\partial h^i_{.\mu}} = (\frac{\partial(\sqrt{-g}\,L_G)}{\partial h^i_{.\mu}})_\Gamma + \frac{\partial(\sqrt{-g}\,L_G)}{\partial \Gamma^{ab}_{..\alpha}} \frac{\partial \Gamma^{ab}_{..\alpha}}{\partial h^i_{.\mu}} + \frac{\partial(\sqrt{-g}\,L_G)}{\partial \Gamma^{ab}_{..\alpha,\beta}} \frac{\partial \Gamma^{ab}_{..\alpha,\beta}}{\partial h^i_{.\mu}} \qquad (39)$$

$$\frac{\partial(\sqrt{-g}\,L_G^*)}{\partial h^i_{.\mu,\lambda}} = (\frac{\partial(\sqrt{-g}\,L_G)}{\partial h^i_{.\mu,\lambda}})_\Gamma + \frac{\partial(\sqrt{-g}\,L_G)}{\partial \Gamma^{ab}_{..\alpha}} \frac{\partial \Gamma^{ab}_{..\alpha}}{\partial h^i_{.\mu,\lambda}} + \frac{\partial(\sqrt{-g}\,L_G)}{\partial \Gamma^{ab}_{..\alpha,\beta}} \frac{\partial \Gamma^{ab}_{..\alpha,\beta}}{\partial h^i_{.\mu,\lambda}} \qquad (40)$$

$$\frac{\partial(\sqrt{-g}\,L_G^*)}{\partial h^i_{.\mu,\lambda\sigma}} = \frac{\partial(\sqrt{-g}\,L_G)}{\partial \Gamma^{ab}_{..\alpha,\beta}} \frac{\partial \Gamma^{ab}_{..\alpha,\beta}}{\partial h^i_{.\mu,\lambda\sigma}} \qquad (41)$$

On the other hand it is evident that $L_M^*$ and $L_G^*$ relating to Eqs.(12,13) satisfy also

$\delta_0(\sqrt{-g}\,\Lambda) + (\xi^\mu \sqrt{-g}\,\Lambda)_{,\mu} \equiv 0$, where $\Lambda = L_M^*$ or $\Lambda = L_G^*$ or $\Lambda = L_M^* + L_G^*$. Owing to the independent arbitrariness of $\varepsilon^{mn}_{,}(x)$, $\varepsilon^{mn}_{,\lambda}(x)$, $\varepsilon^{mn}_{,\lambda\sigma}(x)$, $\xi^\alpha(x)$, $\xi^\alpha_{,\mu}(x)$, $\xi^\alpha_{,\mu\lambda}(x)$ and $\xi^\alpha_{,\mu\lambda\sigma}(x)$, we obtain another set of identities [9]  (if $\Lambda = L_G^*$, $\frac{\partial(\sqrt{-g}\,\Lambda)}{\partial \psi} = 0$, $\frac{\partial(\sqrt{-g}\,\Lambda)}{\partial \psi_{,\lambda}} = 0$ ):

$$\frac{1}{2}\frac{\partial(\sqrt{-g}\,\Lambda)}{\partial \psi}\sigma_{mn}\psi + \frac{1}{2}\frac{\partial(\sqrt{-g}\,\Lambda)}{\partial \psi_{,\lambda}}\sigma_{mn}\psi_{,\lambda} + \frac{\partial(\sqrt{-g}\,\Lambda)}{\partial h^{m}_{.\mu}}h_{n\mu}$$
$$+ \frac{\partial(\sqrt{-g}\,\Lambda)}{\partial h^{m}_{.\mu,\lambda}}h_{n\mu,\lambda} + \frac{\partial(\sqrt{-g}\,\Lambda)}{\partial h^{m}_{.\mu,\lambda\sigma}}h_{n\mu,\lambda\sigma} = 0 \qquad (42)$$

$$\frac{1}{2}\frac{\partial(\sqrt{-g}\,\Lambda)}{\partial \psi_{,\lambda}}\sigma_{mn}\psi + \frac{\partial(\sqrt{-g}\,\Lambda)}{\partial h^{m}_{.\mu,\lambda}}h_{n\mu} + 2\frac{\partial(\sqrt{-g}\,\Lambda)}{\partial h^{m}_{.\mu,\lambda\sigma}}h_{n\mu,\sigma} = 0 \qquad (43)$$

$$\frac{\partial(\sqrt{-g}\,\Lambda)}{\partial h^{m}_{.\mu,\lambda\sigma}}h_{n\mu} = \frac{\partial(\sqrt{-g}\,\Lambda)}{\partial h^{n}_{.\mu,\lambda\sigma}}h_{m\mu} \qquad (44)$$

$$\frac{\partial(\sqrt{-g}\,\Lambda)}{\partial \psi}\psi_{,\alpha} + \frac{\partial(\sqrt{-g}\,\Lambda)}{\partial \psi_{,\lambda}}\psi_{,\lambda\alpha} + \frac{\partial(\sqrt{-g}\,\Lambda)}{\partial h^{i}_{.\mu}}h^{i}_{.\mu,\alpha}$$
$$+ \frac{\partial(\sqrt{-g}\,\Lambda)}{\partial h^{i}_{.\mu,\lambda}}h^{i}_{.\mu,\lambda\alpha} + \frac{\partial(\sqrt{-g}\,\Lambda)}{\partial h^{i}_{.\mu,\lambda\sigma}}h^{i}_{.\mu,\lambda\sigma\alpha} - (\sqrt{-g}\,\Lambda)_{,\alpha} = 0 \qquad (45)$$

$$\frac{\partial(\sqrt{-g}\,\Lambda)}{\partial \psi_{,\lambda}}\psi_{,\alpha} + \frac{\partial(\sqrt{-g}\,\Lambda)}{\partial h^{i}_{.\lambda}}h^{i}_{.\alpha} + \frac{\partial(\sqrt{-g}\,\Lambda)}{\partial h^{i}_{.\mu,\lambda}}h^{i}_{.\mu,\alpha} + \frac{\partial(\sqrt{-g}\,\Lambda)}{\partial h^{i}_{.\lambda,\mu}}h^{i}_{.\alpha,\mu}$$
$$+ \frac{\partial(\sqrt{-g}\,\Lambda)}{\partial h^{i}_{.\lambda,\mu\sigma}}h^{i}_{.\alpha,\mu\sigma} + 2\frac{\partial(\sqrt{-g}\,\Lambda)}{\partial h^{i}_{.\mu,\lambda\sigma}}h^{i}_{.\mu,\sigma\alpha} - \sqrt{-g}\,\Lambda\,\delta^{\lambda}_{\alpha} = 0 \qquad (46)$$

$$\frac{\partial(\sqrt{-g}\,\Lambda)}{\partial h^i_{.\mu,\lambda}}h^i_{.\alpha} + \frac{\partial(\sqrt{-g}\,\Lambda)}{\partial h^i_{.\mu,\lambda\sigma}}h^i_{.\alpha,\sigma} + \frac{\partial(\sqrt{-g}\,\Lambda)}{\partial h^i_{.\sigma,\lambda\mu}}h^i_{.\sigma,\alpha} - \frac{\partial}{\partial x^\sigma}\left(\frac{\partial(\sqrt{-g}\,\Lambda)}{\partial h^i_{.\mu,\lambda\sigma}}\right)h^i_{.\alpha}$$

$$= -\frac{\partial}{\partial x^\sigma}\left(\frac{\partial(\sqrt{-g}\,\Lambda)}{\partial h^i_{.\mu,\lambda\sigma}}h^i_{.\alpha}\right) \tag{47}$$

$$\frac{\partial(\sqrt{-g}\,\Lambda)}{\partial h^i_{.\mu,\lambda\sigma}}h^i_{.\alpha} + \frac{\partial(\sqrt{-g}\,\Lambda)}{\partial h^i_{.\lambda,\sigma\mu}}h^i_{.\alpha} + \frac{\partial(\sqrt{-g}\,\Lambda)}{\partial h^i_{.\sigma,\mu\lambda}}h^i_{.\alpha} = 0 \tag{48}$$

In addition there are the relations:

$$\frac{\partial(\sqrt{-g}\Lambda)}{\partial \Gamma^{ab}_{..\alpha}}\delta_0\Gamma^{ab}_{..\alpha} = \frac{\partial(\sqrt{-g}\Lambda)}{\partial \Gamma^{ab}_{..\alpha}}\left(\frac{\partial \Gamma^{ab}_{..\alpha}}{\partial h^i_{.\mu}}\delta_0 h^i_{.\mu} + \frac{\partial \Gamma^{ab}_{..\alpha}}{\partial h^i_{.\mu,\lambda}}\delta_0 h^i_{.\mu,\lambda}\right) \tag{49}$$

$$\frac{\partial(\sqrt{-g}\Lambda)}{\partial \Gamma^{ab}_{..\alpha,\beta}}\delta_0\Gamma^{ab}_{..\alpha,\beta} = \frac{\partial(\sqrt{-g}\Lambda)}{\partial \Gamma^{ab}_{..\alpha,\beta}}\left(\frac{\partial \Gamma^{ab}_{..\alpha,\beta}}{\partial h^i_{.\mu}}\delta_0 h^i_{.\mu} + \frac{\partial \Gamma^{ab}_{..\alpha,\beta}}{\partial h^i_{.\mu,\lambda}}\delta_0 h^i_{.\mu,\lambda}\right.$$

$$\left. + \frac{\partial \Gamma^{ab}_{..\alpha,\beta}}{\partial h^i_{.\mu,\lambda\sigma}}\delta_0 h^i_{.\mu,\lambda\sigma}\right) \tag{50}$$

Utilizing these relations and those in Eqs.(33-37,39-41), and carrying out some complicated calculations, it can be proven that the identities Eqs.(25-30) are equivalent to the identities Eqs.(42-48).

### 3. Possible forms of the Lagrangians under the symmetry of transformations Eqs.(20,21)

In this section we will prove that, due to the requirement of the action integrals of a gravitational system being invariant under the transformations Eqs.(20,21), the possible forms of the Lagrangian densities Eq.(10) and Eq.(11) might be:

$$\sqrt{-g(x)}\,L_M(x) = \sqrt{-g(x)}\,L^{\#}_M[\psi(x);\psi_{|\mu}(x);R^{ij}_{..\mu\nu}(x)\,;T^i_{.\mu\nu}(x)\,;h^i_{.\mu}(x)\,] \tag{16}$$

and $\quad \sqrt{-g}\,L_G(x) = \sqrt{-g}\,L^{\#}_G[R^{ij}_{..\mu\nu}(x)\,;T^i_{.\mu\nu}(x)\,;h^i_{.\mu}(x)\,] \tag{17}$

respectively. The proof of Eq.(16) is given in the following:

Eq.(27) means that $\Gamma^{ij}_{..\mu,\nu}(x)$ appears in $\sqrt{-g}\,L_M(x)$ only through a curvature tensor field $R^{ij}_{..\mu\nu}(x)$

because $2\frac{\partial(\sqrt{-g}\,L_M)}{\partial R^{ij}_{..\nu\mu}} \equiv \frac{\partial(\sqrt{-g}\,L_M)}{\partial \Gamma^{ij}_{..\mu,\nu}}$ ; Eq.(31) means that $h^i_{.\mu,\nu}(x)$ appears in $\sqrt{-g}\,L_M(x)$ only through torsion tensor field $T^i_{.\mu\nu}(x)$ because $\frac{\partial(\sqrt{-g}\,L_M)}{\partial T^i_{.\mu\nu}} \equiv \frac{\partial(\sqrt{-g}\,L_M)}{\partial h^i_{.\mu,\nu}}$ ; Eq.(26) means that $\Gamma^{mn}_{..\nu}(x)$ appears in

$\sqrt{-g}\,L_M(x)$ only through covariant derivative $\psi_{|\mu}(x)$ and curvature tensor field $R^{ij}_{..\mu\nu}(x)$ and torsion

tensor field $T^i_{.\mu\nu}(x)$ because

$$\frac{\partial(\sqrt{-g}\,L_M)}{\partial \Gamma^{mn}_{..\lambda}} = \frac{1}{2}\frac{\partial(\sqrt{-g}\,L_M)}{\partial \psi_{,\lambda}}\sigma_{mn}\psi + \frac{\partial(\sqrt{-g}\,L_M)}{\partial h^m_{.\mu,\lambda}}h_{n\mu} + 2\frac{\partial(\sqrt{-g}\,L_M)}{\partial \Gamma^{km}_{..\mu,\lambda}}\Gamma^k_{.n\mu}$$

$$= \frac{\partial(\sqrt{-g}\,L_M)}{\partial \psi_{|\mu}}\frac{\partial \psi_{|\mu}}{\partial \Gamma^{mn}_{..\lambda}} + \frac{\partial(\sqrt{-g}\,L_M)}{\partial T^i_{.\alpha\beta}}\frac{\partial T^i_{.\alpha\beta}}{\partial \Gamma^{mn}_{..\lambda}} + \frac{\partial(\sqrt{-g}\,L_M)}{\partial R^{ij}_{..\alpha\beta}}\frac{\partial R^{ij}_{..\alpha\beta}}{\partial \Gamma^{mn}_{..\lambda}}$$

Hence the matter Lagrangian density $\sqrt{-g}\,L_M(x)$ should take the form denoted by Eq.(16).

On the other hand if there exists the relation Eq.(16), we must have:

$$\sqrt{-g(x)}\,L_M(x) = \sqrt{-g(x)}\,L^{\#}_M(x) = \sqrt{-g(x)}\,L^{\#}_M[\,\psi(x);\psi_{|\alpha}[\psi(x);\psi_{,\mu}(x)\,;\Gamma^{ij}_{..\mu}(x)];$$

$$R^{ab}_{..\alpha\beta}[\Gamma^{ij}_{..\mu}(x)\,;\Gamma^{ij}_{..\mu,\lambda}(x)];T^a_{.\alpha\beta}[h^i_{.\mu}(x);h^i_{.\mu.\lambda}(x);\Gamma^{ij}_{..\mu}(x)];h^i_{.\mu}(x)\,] \qquad (51)$$

$$= \sqrt{-g(x)}\,L_M[\,\psi(x);\psi_{,\lambda}(x);h^i_{.\mu}(x)\,;h^i_{.\mu,\lambda}(x);\Gamma^{ij}_{..\mu}(x);\Gamma^{ij}_{..\mu,\lambda}(x)\,]$$

Therefore from Eq.(51)

$$\delta_0(\sqrt{-g}\,L_M^{\#}) = \frac{\partial(\sqrt{-g}\,L_M^{\#})}{\partial \psi}\delta_0\psi + \frac{\partial(\sqrt{-g}\,L_M^{\#})}{\partial \psi_{|\alpha}}\left(\frac{\partial \psi_{|\alpha}}{\partial \psi}\delta_0\psi + \frac{\partial \psi_{|\alpha}}{\partial \psi_{,\mu}}\delta_0\psi_{,\mu} + \frac{\partial \psi_{|\alpha}}{\partial \Gamma^{ij}_{..\mu}}\delta_0\Gamma^{ij}_{..\mu}\right)$$

$$+ \frac{\partial(\sqrt{-g}\,L_M^{\#})}{\partial T^{a}_{.\alpha\beta}}\left(\frac{\partial T^{a}_{.\alpha\beta}}{\partial h^{i}_{.\mu}}\delta_0 h^{i}_{.\mu} + \frac{\partial T^{a}_{.\alpha\beta}}{\partial h^{i}_{.\mu,\lambda}}\delta_0 h^{i}_{.\mu,\lambda} + \frac{\partial T^{a}_{.\alpha\beta}}{\partial \Gamma^{ij}_{.\mu}}\delta_0\Gamma^{ij}_{.\mu}\right)$$

$$+ \frac{\partial(\sqrt{-g}\,L_M^{\#})}{\partial R^{ab}_{..\alpha\beta}}\left(\frac{\partial R^{ab}_{..\alpha\beta}}{\partial \Gamma^{ij}_{..\mu}}\delta_0\Gamma^{ij}_{..\mu} + \frac{\partial R^{ab}_{..\alpha\beta}}{\partial \Gamma^{ij}_{..\mu,\lambda}}\delta_0\Gamma^{ij}_{..\mu,\lambda}\right) + \frac{\partial(\sqrt{-g}\,L_M^{\#})}{\partial h^{i}_{.\mu}}\delta_0 h^{i}_{.\mu}$$

$$= \delta_0(\sqrt{-g}\,L_M) = \frac{\partial(\sqrt{-g}\,L_M)}{\partial \psi}\delta_0\psi + \frac{\partial(\sqrt{-g}\,L_M)}{\partial \psi_{,\lambda}}\delta_0\psi_{,\lambda} + \frac{\partial(\sqrt{-g}\,L_M)}{\partial h^{i}_{.\mu}}\delta_0 h^{i}_{.\mu}$$

$$+ \frac{\partial(\sqrt{-g}\,L_M)}{\partial h^{i}_{.\mu,\lambda}}\delta_0 h^{i}_{.\mu,\lambda} + \frac{\partial(\sqrt{-g}\,L_M)}{\partial \Gamma^{ij}_{..\mu}}\delta_0\Gamma^{ij}_{..\mu} + \frac{\partial(\sqrt{-g}\,L_M)}{\partial \Gamma^{ij}_{..\mu,\lambda}}\delta_0\Gamma^{ij}_{..\mu,\lambda}$$

(52)

Thus we have

$$\frac{\partial(\sqrt{-g}\,L_M)}{\partial \psi} = \frac{\partial(\sqrt{-g}\,L_M^{\#})}{\partial \psi} + \frac{\partial(\sqrt{-g}\,L_M^{\#})}{\partial \psi_{|\alpha}}\frac{\partial \psi_{|\alpha}}{\partial \psi}\ , \tag{53}$$

$$\frac{\partial(\sqrt{-g}\,L_M)}{\partial \psi_{,\lambda}} = \frac{\partial(\sqrt{-g}\,L_M^{\#})}{\partial \psi_{|\alpha}}\frac{\partial \psi_{|\alpha}}{\partial \psi_{,\lambda}} \tag{54}$$

$$\frac{\partial(\sqrt{-g}\,L_M)}{\partial h^{i}_{.\mu}} = \frac{\partial(\sqrt{-g}\,L_M^{\#})}{\partial h^{i}_{.\mu}} + \frac{\partial(\sqrt{-g}\,L_M^{\#})}{\partial T^{a}_{.\alpha\beta}}\frac{\partial T^{a}_{.\alpha\beta}}{\partial h^{i}_{.\mu}} \tag{55}$$

$$\frac{\partial(\sqrt{-g}\,L_M)}{\partial h^{i}_{.\mu,\lambda}} = \frac{\partial(\sqrt{-g}\,L_M^{\#})}{\partial T^{a}_{.\alpha\beta}}\frac{\partial T^{a}_{.\alpha\beta}}{\partial h^{i}_{.\mu,\lambda}} \tag{56}$$

$$\frac{\partial(\sqrt{-g}\,L_M)}{\partial \Gamma^{ij}_{..\mu}} = \frac{\partial(\sqrt{-g}\,L_M^{\#})}{\partial \psi_{|\alpha}}\frac{\partial \psi_{|\alpha}}{\partial \Gamma^{ij}_{..\mu}} + \frac{\partial(\sqrt{-g}\,L_M^{\#})}{\partial T^{a}_{.\alpha\beta}}\frac{\partial T^{a}_{.\alpha\beta}}{\partial \Gamma^{ij}_{..\mu}}$$

$$+ \frac{\partial(\sqrt{-g}\,L_M^{\#})}{\partial R^{ab}_{..\alpha\beta}}\frac{\partial R^{ab}_{..\alpha\beta}}{\partial \Gamma^{ij}_{..\mu}} \tag{57}$$

$$\frac{\partial(\sqrt{-g}\,L_M)}{\partial \Gamma^{ij}_{..\mu,\lambda}} = \frac{\partial(\sqrt{-g}\,L_M^{\#})}{\partial R^{ab}_{..\alpha\beta}}\frac{\partial R^{ab}_{..\alpha\beta}}{\partial \Gamma^{ij}_{..\mu,\lambda}} \tag{58}$$

Using Eqs.(53-58) and Eqs.(25-31) we have the following identities:

$$\frac{1}{2}\left(\frac{\partial(\sqrt{-g}\,L_M^{\#})}{\partial \psi} + \frac{\partial(\sqrt{-g}\,L_M^{\#})}{\partial \psi_{|\alpha}}\frac{\partial \psi_{|\alpha}}{\partial \psi}\right)\sigma_{mn}\psi + \frac{1}{2}\frac{\partial(\sqrt{-g}\,L_M^{\#})}{\partial \psi_{|\alpha}}\frac{\partial \psi_{|\alpha}}{\partial \psi_{,\lambda}}\sigma_{mn}\psi_{,\lambda}$$

$$+\left(\frac{\partial(\sqrt{-g}\,L_M^{\#})}{\partial h^{m}_{.\mu}} + \frac{\partial(\sqrt{-g}\,L_M^{\#})}{\partial T^{a}_{.\alpha\beta}}\frac{\partial T^{a}_{.\alpha\beta}}{\partial h^{m}_{.\mu}}\right)h_{n\mu} + \frac{\partial(\sqrt{-g}\,L_M^{\#})}{\partial T^{a}_{.\alpha\beta}}\frac{\partial T^{a}_{.\alpha\beta}}{\partial h^{m}_{.\mu,\lambda}}h_{n\mu,\lambda}$$

$$+ 2\left(\frac{\partial \sqrt{-g}\,L_M^{\#}}{\partial \psi_{|\alpha}}\frac{\partial \psi_{|\alpha}}{\partial \Gamma^{km}_{..\mu}} + \frac{\partial(\sqrt{-g}\,L_M^{\#})}{\partial T^{a}_{.\alpha\beta}}\frac{\partial T^{a}_{.\alpha\beta}}{\partial \Gamma^{km}_{..\mu}} + \frac{\partial(\sqrt{-g}\,L_M^{\#})}{\partial R^{ab}_{..\alpha\beta}}\frac{\partial R^{ab}_{..\alpha\beta}}{\partial \Gamma^{km}_{..\mu}}\right)\Gamma^{k}_{.n\mu} \tag{59}$$

$$+ 2\frac{\partial(\sqrt{-g}\,L_M^{\#})}{\partial R^{ab}_{..\alpha\beta}}\frac{\partial R^{ab}_{..\alpha\beta}}{\partial \Gamma^{km}_{..\mu,\lambda}}\Gamma^{k}_{.n\mu,\lambda} = 0$$

$$\frac{1}{2}\frac{\partial(\sqrt{-g}\,L_M^{\#})}{\partial \psi_{|\alpha}}\frac{\partial \psi_{|\alpha}}{\partial \psi_{,\lambda}}\sigma_{mn}\psi + \frac{\partial(\sqrt{-g}\,L_M^{\#})}{\partial T^{a}_{.\alpha\beta}}\frac{\partial T^{a}_{.\alpha\beta}}{\partial h^{m}_{.\mu,\lambda}}h_{n\mu} + 2\frac{\partial(\sqrt{-g}\,L_M^{\#})}{\partial R^{ab}_{..\alpha\beta}}\frac{\partial R^{ab}_{..\alpha\beta}}{\partial \Gamma^{km}_{..\mu,\lambda}}\Gamma^{k}_{.n\mu}$$

$$= \frac{\partial(\sqrt{-g}\,L_M^{\#})}{\partial \psi_{|\alpha}}\frac{\partial \psi_{|\alpha}}{\partial \Gamma^{mn}_{..\lambda}} + \frac{\partial(\sqrt{-g}\,L_M^{\#})}{\partial T^{a}_{.\alpha\beta}}\frac{\partial T^{a}_{.\alpha\beta}}{\partial \Gamma^{mn}_{..\lambda}} + \frac{\partial(\sqrt{-g}\,L_M^{\#})}{\partial R^{ab}_{..\alpha\beta}}\frac{\partial R^{ab}_{..\alpha\beta}}{\partial \Gamma^{mn}_{..\lambda}} \tag{60}$$

$$\frac{\partial(\sqrt{-g}\,L_M^{\#})}{\partial R^{ab}_{..\alpha\beta}}\frac{\partial R^{ab}_{..\alpha\beta}}{\partial \Gamma^{mn}_{..\mu,\nu}} = -\frac{\partial(\sqrt{-g}\,L_M^{\#})}{\partial R^{ab}_{..\alpha\beta}}\frac{\partial R^{ab}_{..\alpha\beta}}{\partial \Gamma^{mn}_{..\nu,\mu}} \tag{61}$$

$$\left(\frac{\partial(\sqrt{-g}\,L_M^{\#})}{\partial \psi} + \frac{\partial(\sqrt{-g}\,L_M^{\#})}{\partial \psi_{|\beta}}\frac{\partial \psi_{|\beta}}{\partial \psi}\right)\psi_{,\alpha} + \frac{\partial(\sqrt{-g}\,L_M^{\#})}{\partial \psi_{|\beta}}\frac{\partial \psi_{|\beta}}{\partial \psi_{,\lambda}}\psi_{,\lambda\alpha}$$

$$+\left(\frac{\partial(\sqrt{-g}\,L_M^{\#})}{\partial h^i_{.\mu}} + \frac{\partial(\sqrt{-g}\,L_M^{\#})}{\partial T^a_{.\beta\sigma}}\frac{\partial T^a_{.\beta\sigma}}{\partial h^i_{.\mu}}\right)h^i_{.\mu,\alpha} + \frac{\partial(\sqrt{-g}\,L_M^{\#})}{\partial T^a_{.\beta\sigma}}\frac{\partial T^a_{.\beta\sigma}}{\partial h^i_{.\mu,\lambda}}h^i_{.\mu,\alpha\lambda}$$

$$+\left(\frac{\partial(\sqrt{-g}\,L_M^{\#})}{\partial \psi_{|\beta}}\frac{\partial \psi_{|\beta}}{\partial \Gamma^{ij}_{..\mu}} + \frac{\partial(\sqrt{-g}\,L_M^{\#})}{\partial T^a_{.\beta\sigma}}\frac{\partial T^a_{.\beta\sigma}}{\partial \Gamma^{ij}_{..\mu}} + \frac{\partial(\sqrt{-g}\,L_M^{\#})}{\partial R^{ab}_{..\beta\sigma}}\frac{\partial R^{ab}_{..\beta\sigma}}{\partial \Gamma^{ij}_{..\mu}}\right)\Gamma^{ij}_{..\mu,\alpha}$$

$$+\frac{\partial(\sqrt{-g}\,L_M^{\#})}{\partial R^{ab}_{..\beta\sigma}}\frac{\partial R^{ab}_{..\beta\sigma}}{\partial \Gamma^{ij}_{..\mu,\lambda}}\Gamma^{ij}_{..\mu,\alpha\lambda} - (\sqrt{-g}\,L_M^{\#})_{,\alpha} = 0 \tag{62}$$

$$\frac{\partial(\sqrt{-g}\,L_M^{\#})}{\partial \psi_{|\rho}}\frac{\partial \psi_{|\rho}}{\partial \psi_{,\lambda}}\psi_{,\alpha} + \left(\frac{\partial(\sqrt{-g}\,L_M^{\#})}{\partial h^i_{.\lambda}} + \frac{\partial(\sqrt{-g}\,L_M^{\#})}{\partial T^a_{.\beta\sigma}}\frac{\partial T^a_{.\beta\sigma}}{\partial h^i_{.\lambda}}\right)h^i_{.\alpha}$$

$$+\frac{\partial(\sqrt{-g}\,L_M^{\#})}{\partial T^a_{.\beta\sigma}}\frac{\partial T^a_{.\beta\sigma}}{\partial h^i_{.\mu,\lambda}}(h^i_{.\mu,\alpha} - h^i_{.\alpha,\mu})$$

$$+\left(\frac{\partial(\sqrt{-g}\,L_M^{\#})}{\partial \psi_{|\rho}}\frac{\partial \psi_{|\rho}}{\partial \Gamma^{ij}_{..\lambda}} + \frac{\partial(\sqrt{-g}\,L_M^{\#})}{\partial T^a_{.\beta\sigma}}\frac{\partial T^a_{.\beta\sigma}}{\partial \Gamma^{ij}_{..\lambda}} + \frac{\partial(\sqrt{-g}\,L_M^{\#})}{\partial R^{ab}_{..\beta\sigma}}\frac{\partial R^{ab}_{..\beta\sigma}}{\partial \Gamma^{ij}_{..\lambda}}\right)\Gamma^{ij}_{..\alpha}$$

$$+\frac{\partial(\sqrt{-g}\,L_M^{\#})}{\partial R^{ab}_{..\beta\sigma}}\frac{\partial R^{ab}_{..\beta\sigma}}{\partial \Gamma^{ij}_{..\mu,\lambda}}(\Gamma^{ij}_{..\mu,\alpha} - \Gamma^{ij}_{..\alpha,\mu}) - \sqrt{-g}\,L_M^{\#}\,\delta^\lambda_\alpha = 0 \tag{63}$$

$$\frac{\partial(\sqrt{-g}\,L_M^{\#})}{\partial T_{.\beta\sigma}^{a}}\frac{\partial T_{.\beta\sigma}^{a}}{\partial h_{.\mu,\lambda}^{i}}h_{.\alpha}^{i}+\frac{\partial(\sqrt{-g}\,L_M^{\#})}{\partial R_{..\beta\sigma}^{ab}}\frac{\partial R_{..\beta\sigma}^{ab}}{\partial \Gamma_{..\mu,\lambda}^{ij}}\Gamma_{..\alpha}^{ij}=0 \tag{64}$$

$$\frac{\partial(\sqrt{-g}\,L_M^{\#})}{\partial T_{.\alpha\beta}^{a}}\frac{\partial T_{.\alpha\beta}^{a}}{\partial h_{.\mu,\lambda}^{i}}=-\frac{\partial(\sqrt{-g}\,L_M^{\#})}{\partial T_{.\alpha\beta}^{a}}\frac{\partial T_{.\alpha\beta}^{a}}{\partial h_{.\lambda,\mu}^{i}} \tag{65}$$

On the other hand, from Eq.(16) we also have:

$$\delta_0(\sqrt{-g}\,L_M^{\#}) = \frac{\partial(\sqrt{-g}\,L_M^{\#})}{\partial \psi}\delta_0\psi + \frac{\partial(\sqrt{-g}\,L_M^{\#})}{\partial \psi_{|\lambda}}\delta_0\psi_{|\lambda} + \frac{\partial(\sqrt{-g}\,L_M^{\#})}{\partial h_{.\mu}^{i}}\delta_0 h_{.\mu}^{i}$$
$$+ \frac{\partial(\sqrt{-g}\,L_M^{\#})}{\partial R_{..\alpha\beta}^{ab}}\delta_0 R_{..\alpha\beta}^{ab} + \frac{\partial(\sqrt{-g}\,L_M^{\#})}{\partial T_{.\alpha\beta}^{a}}\delta_0 T_{.\alpha\beta}^{a} \tag{66}$$

where

$$\delta_0\psi_{|\lambda} = \delta_0\psi_{,\lambda} + \frac{1}{2}\Gamma_{..\lambda}^{mn}\sigma_{mn}\delta_0\psi + \frac{1}{2}(\delta_0\Gamma_{..\lambda}^{mn})\sigma_{mn}\psi \tag{67}$$

$$\delta_0 R_{..\alpha\beta}^{ab} = \delta_0\Gamma_{..\beta,\alpha}^{ab} - \delta_0\Gamma_{..\alpha,\beta}^{ab} + \delta_m^a \eta_{nk}((\delta_0\Gamma_{..\alpha}^{mn})\Gamma_{..\beta}^{kb}$$
$$+ \Gamma_{..\alpha}^{mn}(\delta_0\Gamma_{..\beta}^{kb}) - (\delta_0\Gamma_{..\beta}^{mn})\Gamma_{..\alpha}^{kb} - \Gamma_{..\beta}^{mn}(\delta_0\Gamma_{..\alpha}^{kb})) \tag{68}$$

$$\delta_0 T_{.\alpha\beta}^{a} = \frac{1}{2}\{\delta_0 h_{.\alpha,\beta}^{a} - \delta_0 h_{.\beta,\alpha}^{a} + \delta_m^a \eta_{nk}((\delta_0\Gamma_{..\beta}^{mn})h_{.\alpha}^{k} + \Gamma_{..\beta}^{mn}(\delta_0 h_{.\alpha}^{k})$$
$$- (\delta_0\Gamma_{..\alpha}^{mn})h_{.\beta}^{k} - \Gamma_{..\alpha}^{mn}(\delta_0 h_{.\beta}^{k}))\} \tag{69}$$

Substituting Eqs.(67-69) into Eq.(66) and using

$$\delta_0(\sqrt{-g}\,L_M^{\#}) + (\xi^{\mu}\sqrt{-g}\,L_M^{\#})_{,\mu} \equiv 0;$$ because of the independent arbitrariness of $\varepsilon^{mn}(x)$, $\varepsilon_{,\lambda}^{mn}(x)$, $\varepsilon_{,\lambda\sigma}^{mn}(x)$, $\xi^{\alpha}(x)$, $\xi_{,\mu}^{\alpha}(x)$ and $\xi_{,\mu\lambda}^{\alpha}(x)$, after some lengthy calculations we can obtain the identities Eqs.(59-64) one by one. Hence the identities obtained directly from

$\sqrt{-g} \, L_M^\# \, [\, \psi \,; \psi_{|\alpha} \,; R^{ab}_{..\alpha\beta} \,; T^a_{.\alpha\beta} \,; h^i_{.\mu} \,]$ are just the same as those derived from

$\sqrt{-g} \, L_M^\# \, [\, \psi \,; \psi_{|\alpha} \, [\psi \,; \psi_{,\mu} \,; \Gamma^{ij}_{..\mu}] \,; R^{ab}_{..\alpha\beta} \, [\Gamma^{ij}_{..\mu} \,; \Gamma^{ij}_{..\mu,\lambda}] \,; T^a_{.\alpha\beta} \, [h^i_{.\mu} \,; h^i_{.\mu.\lambda} \,; \Gamma^{ij}_{..\mu}] \,; h^i_{.\mu} \,]$

$= \sqrt{-g} \, L_M \, [\, \psi \,; \psi_{,\lambda} \,; h^i_{.\mu} \,; h^i_{.\mu,\lambda} \,; \Gamma^{ij}_{..\mu} \,; \Gamma^{ij}_{..\mu,\lambda} \,]$

The above analysis prove that the relation

$\sqrt{-g(x)} \, L_M \, (x) = \sqrt{-g(x)} \, L_M \, [\psi(x); \psi_{,\mu}(x); h^i_{.\mu}(x); h^i_{.\mu,\lambda}(x) \,; \Gamma^{ij}_{..\mu}(x) \,; \Gamma^{ij}_{..\mu\lambda}(x) \,]$

$= \sqrt{-g(x)} \, L_M^\#(x) = \sqrt{-g(x)} \, L_M^\# \, [\psi(x); \psi_{|\mu}(x); R^{ij}_{..\mu\nu}(x) \,; T^i_{.\mu\nu}(x) \,; h^i_{.\mu}(x) \,]$

must exist.

With the same method we can also prove the following relations:

$\sqrt{-g(x)} \, L_G \, (x) = \sqrt{-g(x)} \, L_G [h^i_{.\mu}(x); h^i_{.\mu,\lambda}(x) \,; \Gamma^{ij}_{..\mu}(x) \,; \Gamma^{ij}_{..\mu\lambda}(x) \,]$

$= \sqrt{-g(x)} \, L_G^\#(x) = \sqrt{-g(x)} \, L_G^\# [R^{ij}_{..\mu\nu}(x) \,; T^i_{.\mu\nu}(x) \,; h^i_{.\mu}(x) \,]$

Previously we tried to prove Eq.(16) in Ref.[11]. However there appear to be some errors in that paper. In the above analysis we believe we have corrected the flaws.

## 4. Conservation laws for a gravitational system with our further generalized Lagrangian density

Below we shall derive the conservation laws for a gravitational system with our further generalized Lagrangian density denoted by Eqs.(10,11) from the identities Eqs.(25-30) and equations of fields:

$$\frac{\partial(\sqrt{-g} \, L_M)}{\partial \psi} - \frac{\partial}{\partial x^\mu} \frac{\partial(\sqrt{-g} \, L_M)}{\partial \psi_{,\mu}} = 0 \tag{70}$$

$$\frac{\delta(\sqrt{-g} \, L_G)}{\delta h^i_{.\mu}} = \frac{\partial(\sqrt{-g} \, L_G)}{\partial h^i_{.\mu}} - \frac{\partial}{\partial x^\lambda} \frac{\partial(\sqrt{-g} \, L_G)}{\partial h^i_{.\mu,\lambda}}$$

$$= -\frac{\partial(\sqrt{-g} \, L_M)}{\partial h^i_{.\mu}} + \frac{\partial}{\partial x^\lambda} \frac{\partial(\sqrt{-g} \, L_M)}{\partial h^i_{.\mu,\lambda}} = -\frac{\delta(\sqrt{-g} \, L_M)}{\delta h^i_{.\mu}} \tag{71}$$

$$\frac{\delta (\sqrt{-g}\, L_G)}{\delta \Gamma^{ij}_{..\mu}} = \frac{\partial(\sqrt{-g}\, L_G)}{\partial \Gamma^{ij}_{..\mu}} - \frac{\partial}{\partial x^\lambda} \frac{\partial(\sqrt{-g}\, L_G)}{\partial \Gamma^{ij}_{..\mu,\lambda}}$$
$$= -\frac{\partial(\sqrt{-g}\, L_M)}{\partial \Gamma^{ij}_{..\mu}} + \frac{\partial}{\partial x^\lambda} \frac{\partial(\sqrt{-g}\, L_M)}{\partial \Gamma^{ij}_{..\mu,\lambda}} = -\frac{\delta (\sqrt{-g}\, L_M)}{\delta \Gamma^{ij}_{..\mu}}$$

(72)

From them we can get the following relations:

$$\frac{\partial}{\partial x^\lambda}(\sqrt{-g}(L_M + L_G)\delta^\lambda_\alpha - \frac{\partial(\sqrt{-g}\, L_M)}{\partial \psi_{,\lambda}} \psi_{,\alpha} - \frac{\partial(\sqrt{-g}(L_M + L_G))}{\partial h^i_{.\mu,\lambda}} h^i_{.\mu,\alpha}$$
$$- \frac{\partial(\sqrt{-g}(L_M + L_G))}{\partial \Gamma^{ij}_{..\mu,\lambda}} \Gamma^{ij}_{..\mu,\alpha}) = 0$$

(73)

$$\frac{\partial}{\partial x^\lambda}(\frac{1}{2}\frac{\partial(\sqrt{-g}\, L_M)}{\partial \psi_{,\lambda}} \sigma_{mn} \psi + \frac{\partial(\sqrt{-g}(L_M + L_G))}{\partial h^m_{.\mu,\lambda}} h_{n\mu}$$
$$+ 2\frac{\partial(\sqrt{-g}(L_M + L_G))}{\partial \Gamma^{km}_{..\mu,\lambda}} \Gamma^k_{.n\mu}) = 0$$

(74)

$$\sqrt{-g}\, L_M\, \delta^\lambda_\alpha - \frac{\partial(\sqrt{-g}\, L_M)}{\partial \psi_{,\lambda}} \psi_{,\alpha} - \frac{\partial(\sqrt{-g}\, L_M)}{\partial h^i_{.\mu,\lambda}} h^i_{.\mu,\alpha} - \frac{\partial(\sqrt{-g}\, L_M)}{\partial \Gamma^{ij}_{..\mu,\lambda}} \Gamma^{ij}_{..\mu,\alpha}$$
$$= (\frac{\partial(\sqrt{-g}\, L_M)}{\partial h^i_{.\lambda}} - \frac{\partial}{\partial x^\mu}(\frac{\partial(\sqrt{-g}\, L_M)}{\partial h^i_{.\lambda,\mu}}))\, h^i_{.\alpha} + (\frac{\partial(\sqrt{-g}\, L_M)}{\partial \Gamma^{ij}_{..\lambda}} - \frac{\partial}{\partial x^\mu}(\frac{\partial(\sqrt{-g}\, L_M)}{\partial \Gamma^{ij}_{..\lambda,\mu}}))\, \Gamma^{ij}_{..\alpha}$$

(75)

$$\sqrt{-g}\, L_G\, \delta^\lambda_\alpha - \frac{\partial(\sqrt{-g}\, L_G)}{\partial h^i_{.\mu,\lambda}} h^i_{.\mu,\alpha} - \frac{\partial(\sqrt{-g}\, L_G)}{\partial \Gamma^{ij}_{..\mu,\lambda}} \Gamma^{ij}_{..\mu,\alpha}$$
$$= (\frac{\partial(\sqrt{-g}\, L_G)}{\partial h^i_{.\lambda}} - \frac{\partial}{\partial x^\mu}(\frac{\partial(\sqrt{-g}\, L_G)}{\partial h^i_{.\lambda,\mu}}))\, h^i_{.\alpha} + (\frac{\partial(\sqrt{-g}\, L_G)}{\partial \Gamma^{ij}_{..\lambda}} - \frac{\partial}{\partial x^\mu}(\frac{\partial(\sqrt{-g}\, L_G)}{\partial \Gamma^{ij}_{..\lambda,\mu}}))\, \Gamma^{ij}_{..\alpha}$$

(76)

$$\frac{1}{2}\frac{\partial(\sqrt{-g}\,L_M)}{\partial\psi_{,\lambda}}\sigma_{mn}\psi + \frac{\partial(\sqrt{-g}\,L_M)}{\partial h^m_{.\mu,\lambda}}h_{n\mu} + 2\frac{\partial(\sqrt{-g}\,L_M)}{\partial\Gamma^{km}_{..\mu,\lambda}}\Gamma^k_{.n\mu}$$

$$-\frac{\partial}{\partial x^\mu}(\frac{\partial(\sqrt{-g}\,L_M)}{\partial\Gamma^{mn}_{..\lambda,\mu}}) = \frac{\partial(\sqrt{-g}\,L_M)}{\partial\Gamma^{mn}_{..\lambda}} - \frac{\partial}{\partial x^\mu}(\frac{\partial(\sqrt{-g}\,L_M)}{\partial\Gamma^{mn}_{..\lambda,\mu}}) \qquad (77)$$

$$\frac{\partial(\sqrt{-g}\,L_G)}{\partial h^m_{.\mu,\lambda}}h_{n\mu} + 2\frac{\partial(\sqrt{-g}\,L_G)}{\partial\Gamma^{km}_{..\mu,\lambda}}\Gamma^k_{.n\mu}$$

$$-\frac{\partial}{\partial x^\mu}(\frac{\partial(\sqrt{-g}\,L_G)}{\partial\Gamma^{mn}_{..\lambda,\mu}}) = \frac{\partial(\sqrt{-g}\,L_G)}{\partial\Gamma^{mn}_{..\lambda}} - \frac{\partial}{\partial x^\mu}(\frac{\partial(\sqrt{-g}\,L_G)}{\partial\Gamma^{mn}_{..\lambda,\mu}}) \qquad (78)$$

Eq.(73) might be regarded as conservation laws of energy-momentum tensor density for the gravitational system:

$$\frac{\partial}{\partial x^\lambda}(\sqrt{-g}\,t^\lambda_{(M)\alpha} + \sqrt{-g}\,t^\lambda_{(G)\alpha}) = 0 \qquad (79)$$

where

$$\sqrt{-g}\,t^\lambda_{(M)\alpha} = \sqrt{-g}\,L_M\,\delta^\lambda_\alpha - \frac{\partial(\sqrt{-g}\,L_M)}{\partial\psi_{,\lambda}}\psi_{,\alpha}$$

$$-\frac{\partial(\sqrt{-g}\,L_M)}{\partial h^i_{.\mu,\lambda}}h^i_{.\mu,\alpha} - \frac{\partial(\sqrt{-g}\,L_M)}{\partial\Gamma^{ij}_{..\mu,\lambda}}\Gamma^{ij}_{..\mu,\alpha} \qquad (80)$$

and $\quad \sqrt{-g}\,t^\lambda_{(G)\alpha} = \sqrt{-g}\,L_G\,\delta^\lambda_\alpha - \frac{\partial(\sqrt{-g}\,L_G)}{\partial h^i_{.\mu,\lambda}}h^i_{.\mu,\alpha} - \frac{\partial(\sqrt{-g}\,L_G)}{\partial\Gamma^{ij}_{..\mu,\lambda}}\Gamma^{ij}_{..\mu,\alpha} \qquad (81)$

might be interpreted as the energy-momentum tensor density of matter field and of gravitational field respectively. But we must indicated in that $\sqrt{-g}\,t^\lambda_{(M)\alpha}$ and $\sqrt{-g}\,t^\lambda_{(G)\alpha}$ are not tensor densities and Eq.(79) lacks the invariant character it should have in the theories of relativistic gravitation. However if we use Eq. (75) to define

$$\sqrt{-g}\,T^{\lambda}_{(M)\alpha} = \left(\frac{\partial(\sqrt{-g}\,L_M)}{\partial h^i_{.\lambda}} - \frac{\partial}{\partial x^{\mu}}\left(\frac{\partial(\sqrt{-g}\,L_M)}{\partial h^i_{.\lambda,\mu}}\right)\right)h^i_{.\alpha} = \sqrt{-g}\,L_M\,\delta^{\lambda}_{\alpha} - \frac{\partial(\sqrt{-g}\,L_M)}{\partial \psi_{,\lambda}}\psi_{,\alpha} \quad (82)$$

$$-\frac{\partial(\sqrt{-g}\,L_M)}{\partial h^i_{.\mu,\lambda}}h^i_{.\mu,\alpha} - \frac{\partial(\sqrt{-g}\,L_M)}{\partial \Gamma^{ij}_{..\mu,\lambda}}\Gamma^{ij}_{..\mu,\alpha} - \left(\frac{\partial(\sqrt{-g}\,L_M)}{\partial \Gamma^{ij}_{..\lambda}} - \frac{\partial}{\partial x^{\mu}}\left(\frac{\partial(\sqrt{-g}\,L_M)}{\partial \Gamma^{ij}_{..\lambda,\mu}}\right)\right)\Gamma^{ij}_{..\alpha}$$

and use Eq. (76) to define

$$\sqrt{-g}\,T^{\lambda}_{(G)\alpha} = \left(\frac{\partial(\sqrt{-g}\,L_G)}{\partial h^i_{.\lambda}} - \frac{\partial}{\partial x^{\mu}}\left(\frac{\partial(\sqrt{-g}\,L_G)}{\partial h^i_{.\lambda,\mu}}\right)\right)h^i_{.\alpha} = \sqrt{-g}\,L_G\,\delta^{\lambda}_{\alpha} - \frac{\partial(\sqrt{-g}\,L_G)}{\partial h^i_{.\mu,\lambda}}h^i_{.\mu,\alpha}$$

$$-\frac{\partial(\sqrt{-g}\,L_G)}{\partial \Gamma^{ij}_{..\mu,\lambda}}\Gamma^{ij}_{..\mu,\alpha} - \left(\frac{\partial(\sqrt{-g}\,L_G)}{\partial \Gamma^{ij}_{..\lambda}} - \frac{\partial}{\partial x^{\mu}}\left(\frac{\partial(\sqrt{-g}\,L_G)}{\partial \Gamma^{ij}_{..\lambda,\mu}}\right)\right)\Gamma^{ij}_{..\alpha} \quad (83)$$

then we get

$$\sqrt{-g}\,T^{\lambda}_{(M)\alpha} + \sqrt{-g}\,T^{\lambda}_{(G)\alpha} = 0 \quad (84)$$

$$\frac{\partial}{\partial x^{\lambda}}\left(\sqrt{-g}\,T^{\lambda}_{(M)\alpha} + \sqrt{-g}\,T^{\lambda}_{(G)\alpha}\right) = 0 \quad (85)$$

At here $\sqrt{-g}\,T^{.\lambda}_{(M)\alpha}$ and $\sqrt{-g}\,T^{.\lambda}_{(G)\alpha}$ are tensor densities and Eq.(85) is a covariant relation. Hence we will take Eqs.(84,85) to be the conservation laws of energy-momentum tensor density for the gravitational system with our further generalized Lagrangian densities. Historically, Einstein had proposed other conservation laws of energy--momentum tensor density for a gravitational system [12]:

$$\frac{\partial}{\partial x^{\lambda}}\left(\sqrt{-g}\,T^{\lambda}_{(M)\alpha} + \sqrt{-g}\,\tilde{t}^{\lambda}_{(G)\alpha}\right) = 0 \quad (86)$$

where $\quad \sqrt{-g}\,\tilde{t}^{\lambda}_{(G)\alpha} = \sqrt{-g}\,T^{\lambda}_{(G)\alpha} - \frac{\partial}{\partial x^{\beta}}u^{\lambda\beta}_{(G)\alpha} , \quad \frac{\partial}{\partial x^{\beta}}u^{\lambda\beta}_{(G)\alpha} = -\frac{\partial}{\partial x^{\beta}}u^{\beta\lambda}_{(G)\alpha}$

The virtues and defects about Eq.(85) and Eq.(86) have been discussed thoroughly in Refs.[9,12]. Because Eqs.(84,85) have more logical basis and rich physical contents, the author believes that the conservation laws in Eqs.(84,85) might be better than Einstein's conservation laws Eq.(86) [9,12]] and could be tested by future experiments and observations.

Eq.(74) might be regarded as conservation laws of spin density for the gravitational system:

$$\frac{\partial}{\partial x^{\lambda}}\left(\sqrt{-g}\,s^{\lambda}_{(M)mn} + \sqrt{-g}\,s^{\lambda}_{(G)mn}\right) = 0 \quad (87)$$

where

$$\sqrt{-g}\, s^{.\lambda}_{(M)mn} = \frac{1}{2}\frac{\partial(\sqrt{-g}\, L_M)}{\partial \psi_{,\lambda}}\sigma_{mn}\psi + \frac{\partial(\sqrt{-g}\, L_M)}{\partial h^m_{.\mu,\lambda}}h_{n\mu} + 2\frac{\partial(\sqrt{-g}\, L_M)}{\partial \Gamma^{km}_{..\mu,\lambda}}\Gamma^k_{.n\mu} \tag{88}$$

and
$$\sqrt{-g}\, s^{.\lambda}_{(G)mn} = \frac{\partial(\sqrt{-g}\, L_G)}{\partial h^m_{.\mu,\lambda}}h_{n\mu} + \frac{\partial(\sqrt{-g}\, L_G)}{\partial \Gamma^{km}_{..\mu,\lambda}}\Gamma^k_{.n\mu} \tag{89}$$

might be interpreted as the spin density of matter field and of gravitational field respectively. But we must indicated in that $\sqrt{-g}\, s^{.\lambda}_{(M)mn}$, $\sqrt{-g}\, s^{.\lambda}_{(G)mn}$ and Eq.(87) lack at all the invariant character it should have in the spirit of general relativity. However if we use Eq. (77) to define

$$\sqrt{-g}\, S^{.\lambda}_{(M)mn} = \frac{\partial(\sqrt{-g}\, L_M)}{\partial \Gamma^{mn}_{..\lambda}} - \frac{\partial}{\partial x^\mu}\left(\frac{\partial(\sqrt{-g}\, L_M)}{\partial \Gamma^{mn}_{..\lambda,\mu}}\right) = \frac{1}{2}\frac{\partial(\sqrt{-g}\, L_M)}{\partial \psi_{,\lambda}}\sigma_{mn}\psi$$
$$+ \frac{\partial(\sqrt{-g}\, L_M)}{\partial h^m_{.\mu,\lambda}}h_{n\mu} + 2\frac{\partial(\sqrt{-g}\, L_M)}{\partial \Gamma^{km}_{..\mu,\lambda}}\Gamma^k_{.n\mu} - \frac{\partial}{\partial x^\mu}\left(\frac{\partial(\sqrt{-g}\, L_M)}{\partial \Gamma^{mn}_{..\lambda,\mu}}\right) \tag{90}$$

and use Eq. (78) to define

$$\sqrt{-g}\, S^{.\lambda}_{(G)mn} = \frac{\partial(\sqrt{-g}\, L_G)}{\partial \Gamma^{mn}_{..\lambda}} - \frac{\partial}{\partial x^\mu}\left(\frac{\partial(\sqrt{-g}\, L_G)}{\partial \Gamma^{mn}_{..\lambda,\mu}}\right)$$
$$= \frac{\partial(\sqrt{-g}\, L_G)}{\partial h^m_{.\mu,\lambda}}h_{n\mu} + 2\frac{\partial(\sqrt{-g}\, L_G)}{\partial \Gamma^{km}_{..\mu,\lambda}}\Gamma^k_{.n\mu} - \frac{\partial}{\partial x^\mu}\left(\frac{\partial(\sqrt{-g}\, L_G)}{\partial \Gamma^{mn}_{..\lambda,\mu}}\right) \tag{91}$$

then we get

$$\sqrt{-g}\, S^{.\lambda}_{(M)mn} + \sqrt{-g}\, S^{.\lambda}_{(G)mn} = 0 \tag{92}$$

$$\frac{\partial}{\partial x^\lambda}\left(\sqrt{-g}\, S^{.\lambda}_{(M)mn} + \sqrt{-g}\, S^{.\lambda}_{(G)mn}\right) = 0 \tag{93}$$

At here $\sqrt{-g}\, S^{.\lambda}_{(M)mn}$, $\sqrt{-g}\, S^{.\lambda}_{(G)mn}$ and Eq.(93) all have the invariant character, hence we will take Eqs.(92,93) to be the conservation laws of spin density for the gravitational system with our further generalized Lagrangian densities.

## 5. Some special cases of Eq.(10) and Eq.(11)

We have indicated that

$$\sqrt{-g(x)}\, L_M(x) = \sqrt{-g(x)}\, L_M[\psi(x); \psi_{,\mu}(x); h^i_{\cdot\mu}(x); h^i_{\cdot\mu,\lambda}(x); \Gamma^{ij}_{\cdot\cdot\mu}(x)]$$

$$\sqrt{-g(x)}\, L_M(x) = \sqrt{-g(x)}\, L_M[\psi(x); \psi_{,\mu}(x); h^i_{\cdot\mu}(x); \Gamma^{ij}_{\cdot\cdot\mu}(x); \Gamma^{ij}_{\cdot\cdot\mu,\lambda}(x)]$$

$$\sqrt{-g(x)}\, L_M(x) = \sqrt{-g(x)}\, L_M[\psi(x); \psi_{,\mu}(x); h^i_{\cdot\mu}(x); \Gamma^{ij}_{\cdot\cdot\mu}(x)]$$

are all the special cases of Eq.(10), and It is evident that

$$\sqrt{-g(x)}\, L_G(x) = \sqrt{-g(x)}\, L_G[h^i_{\cdot\mu}(x); \Gamma^{ij}_{\cdot\cdot\mu}(x); \Gamma^{ij}_{\cdot\cdot\mu,\lambda}(x)]$$

$$\sqrt{-g(x)}\, L_G(x) = \sqrt{-g(x)}\, L_G[h^i_{\cdot\mu}(x); h^i_{\cdot\mu,\lambda}(x)]$$

are all the special cases of Eq.(11). With the same method to prove

$$\sqrt{-g(x)}\, L_M(x) = \sqrt{-g(x)}\, L_M[\psi(x); \psi_{,\mu}(x); h^i_{\cdot\mu}(x); h^i_{\cdot\mu,\lambda}(x); \Gamma^{ij}_{\cdot\cdot\mu}(x); \Gamma^{ij}_{\cdot\cdot\mu\lambda}(x)]$$

$$= \sqrt{-g(x)}\, L^{\#}_M(x) = \sqrt{-g(x)}\, L^{\#}_M[\psi(x); \psi_{|\mu}(x); R^{ij}_{\cdot\cdot\mu\nu}(x); T^i_{\cdot\mu\nu}(x); h^i_{\cdot\mu}(x)]$$

we can also prove the following relations:

$$\sqrt{-g(x)}\, L_M(x) = \sqrt{-g(x)}\, L_M[\psi(x); \psi_{,\mu}(x); h^i_{\cdot\mu}(x); h^i_{\cdot\mu,\lambda}(x); \Gamma^{ij}_{\cdot\cdot\mu}(x)]$$
$$= \sqrt{-g(x)}\, L^{\#}_M(x) = \sqrt{-g(x)}\, L^{\#}_M[\psi(x); \psi_{|\mu}(x); T^i_{\cdot\mu\nu}(x); h^i_{\cdot\mu}(x)] \tag{94}$$

$$\sqrt{-g(x)}\, L_M(x) = \sqrt{-g(x)}\, L_M[\psi(x); \psi_{,\mu}(x); h^i_{\cdot\mu}(x); \Gamma^{ij}_{\cdot\cdot\mu}(x); \Gamma^{ij}_{\cdot\cdot\mu,\lambda}(x)]$$
$$= \sqrt{-g(x)}\, L^{\#}_M(x) = \sqrt{-g(x)}\, L^{\#}_M[\psi(x); \psi_{|\mu}(x); R^{ij}_{\cdot\cdot\mu\nu}(x); h^i_{\cdot\mu}(x)] \tag{95}$$

$$\sqrt{-g(x)}\, L_M(x) = \sqrt{-g(x)}\, L_M[\psi(x); \psi_{,\mu}(x); h^i_{\cdot\mu}(x); \Gamma^{ij}_{\cdot\cdot\mu}(x)]$$
$$= \sqrt{-g(x)}\, L^{\#}_M(x) = \sqrt{-g(x)}\, L^{\#}_M[\psi(x); \psi_{|\mu}(x); h^i_{\cdot\mu}(x)] \tag{96}$$

$$\sqrt{-g(x)}\, L_G(x) = \sqrt{-g(x)}\, L_G[h^i_{.\mu}(x); \Gamma^{ij}_{..\mu}(x); \Gamma^{ij}_{..\mu\lambda}(x)] \qquad (97)$$

$$= \sqrt{-g(x)}\, L^{\#}_G(x) = \sqrt{-g(x)}\, L^{\#}_G[R^{ij}_{..\mu\nu}(x); h^i_{.\mu}(x)]$$

$$\sqrt{-g(x)}\, L_G(x) = \sqrt{-g(x)}\, L_G[h^i_{.\mu}(x); h^i_{.\mu,\lambda}(x)] \qquad (98)$$

$$= \sqrt{-g(x)}\, L^{\#}_G(x) = \sqrt{-g(x)}\, L^{\#}_G[T^i_{.\mu\nu}(x); h^i_{.\mu}(x)]$$

It is not difficult to verify that, for the above special cases of Eq.(10) and Eq.(11), the conservation laws for a gravitational system all have the same mathematical form:

$$\sqrt{-g}\, T^{.\lambda}_{(M)\alpha} + \sqrt{-g}\, T^{.\lambda}_{(G)\alpha} = 0$$

$$\frac{\partial}{\partial x^\lambda}\left(\sqrt{-g}\, T^{.\lambda}_{(M)\alpha} + \sqrt{-g}\, T^{.\lambda}_{(G)\alpha}\right) = 0$$

and

$$\sqrt{-g}\, S^{.\lambda}_{(M)mn} + \sqrt{-g}\, S^{.\lambda}_{(G)mn} = 0$$

$$\frac{\partial}{\partial x^\lambda}\left(\sqrt{-g}\, S^{.\lambda}_{(M)mn} + \sqrt{-g}\, S^{.\lambda}_{(G)mn}\right) = 0$$

Of course the definitions of energy-momentum tensor density and spin density for different Lagrangian densities are different.

From the discussions in the above sections, we have seen that our further generalized Lagrangian density can be used for describing many theories of gravitation. Their general characters are: the gravitational fields could act on the matter field only through covariant derivative, curvature of space-time, and torsion of space-time; the Lagrangian densities of gravitational field are composed of curvature tensor field and torsion tensor field; the conservation laws for a gravitational system all have the same mathematical form. Their peculiarities are: the concrete forms of Lagrangian densities for matter and gravitational field are different, so the couplings between the gravitational fields and matter field are different; the definitions of energy-momentum tensor density and spin density for different Lagrangian densities are different.

**Appendix**
**Ⅰ. Proof of the relation Eq.(5) for the space-time without torsion**

The holonomic connection field $\Gamma^{\mu}_{.\nu\lambda}(x)$ is related to $h^i_{.\mu}(x)$ and $\Gamma^{ij}_{..\mu}(x)$ by [3]

$$\Gamma^{\mu}_{.\nu\lambda}(x) = h^{.\mu}_i(x)[h^i_{.\nu,\lambda}(x) + \Gamma^i_{.j\lambda}(x) h^j_{.\nu}(x)] \qquad (A1)$$

where $\Gamma^i_{.j\lambda}(x) = \eta_{jk}\Gamma^{ik}_{..\lambda}(x)$. In addition, $\Gamma^{ij}_{..\mu}(x) = -\Gamma^{ji}_{..\mu}(x)$. Eq.(A1) can be derived from the requirement:

$$g_{\mu\nu,\lambda} - \Gamma^{\sigma}_{.\mu\lambda}g_{\sigma\nu} - \Gamma^{\sigma}_{.\nu\lambda}g_{\sigma\mu} = 0 \tag{A2}$$

This requirement guarantees that lengths and angles are preserved under parallel displacement [4]. The torsion tensor is defined by [5]

$$T^{\mu}_{.\nu\lambda} = \frac{1}{2}(\Gamma^{\mu}_{.\nu\lambda} - \Gamma^{\mu}_{.\lambda\nu}) \tag{A3}$$

There exists the relation [5]:

$$\Gamma^{\mu}_{.\nu\lambda} = \{^{\mu}_{\nu\lambda}\} + T^{..\mu}_{\nu\lambda} - T^{.\mu}_{\lambda.\nu} + T^{\mu}_{.\nu\lambda} \tag{A4}$$

where
$$\{^{\mu}_{\nu\lambda}\} = \frac{1}{2}g^{\mu\sigma}(g_{\sigma\lambda,\nu} + g_{\sigma\nu,\lambda} - g_{\nu\lambda,\sigma}) \tag{A5}$$

is the Christoffel symbol. Eq.(A4) can be derived from Eqs.(A2,A3,A5). In the space-time without torsion, from Eq.(A4) it is obviously $\Gamma^{\mu}_{.\nu\lambda} = \{^{\mu}_{\nu\lambda}\}$. In this case the relation

$$\begin{aligned}\Gamma^{ij}_{..\mu} &= \frac{1}{2}\eta^{jk}h^{\nu}_k(h^i_{.\mu,\nu} - h^i_{.\nu,\mu}) + \frac{1}{2}\eta^{id}h^{\sigma}_d(h^j_{.\sigma,\mu} - h^j_{.\mu,\sigma}) \\ &+ \frac{1}{2}\eta^{jk}h^{\nu}_k\eta^{id}h^{\sigma}_d\eta_{ab}h^b_{.\mu}(h^a_{.\sigma,\nu} - h^a_{.\nu,\sigma})\end{aligned} \tag{5}$$

can be obtained from Eqs.(A1,A5).

## II. Some useful relations of differential geometry

At here we introduce some useful relations of differential geometry which will be used in this paper.

The curvature tensor related to connection $\{^{\mu}_{\nu\lambda}\}$ is defined by [8]

$$\overset{(\{\})}{R^{\sigma}_{.\lambda\mu\nu}} = \{^{\sigma}_{\lambda\nu}\}_{,\mu} - \{^{\sigma}_{\lambda\mu}\}_{,\nu} + \{^{\sigma}_{\rho\mu}\}\{^{\rho}_{\lambda\nu}\} - \{^{\sigma}_{\rho\nu}\}\{^{\rho}_{\lambda\mu}\} \tag{A6}$$

Similarly the curvature tensor related to connection $\Gamma^{\mu}_{.\nu\lambda}$ is defined by

$$\overset{(\Gamma)}{R^{\sigma}_{.\lambda\mu\nu}} = \Gamma^{\sigma}_{.\lambda\nu,\mu} - \Gamma^{\sigma}_{.\lambda\mu,\nu} + \Gamma^{\sigma}_{.\rho\mu}\Gamma^{\rho}_{.\lambda\nu} - \Gamma^{\sigma}_{.\rho\nu}\Gamma^{\rho}_{.\lambda\mu} \tag{A7}$$

Eq.(A4) suggests that, $\overset{(\Gamma)}{R^{\sigma}_{.\lambda\mu\nu}} \neq \overset{(\{\})}{R^{\sigma}_{.\lambda\mu\nu}}$ for the space-time with torsion; and $\overset{(\Gamma)}{R^{\sigma}_{.\lambda\mu\nu}} = \overset{(\{\})}{R^{\sigma}_{.\lambda\mu\nu}}$ only for the space-time without torsion. We can also define the curvature tensor related to the frame connection $\Gamma^{ij}_{..\mu}$ [3]:

$$R^{ij}_{..\mu\nu} = \Gamma^{ij}_{..\nu,\mu} - \Gamma^{ij}_{..\mu,\nu} + \Gamma^{i}_{.k\mu}\Gamma^{kj}_{..\nu} - \Gamma^{i}_{.k\nu}\Gamma^{kj}_{..\mu} \tag{A8}$$

Using Eq.(A1) it can be verified that $R^{ij}_{..\mu\nu} = \eta^{jk} h^{i}_{.\sigma} h^{\lambda}_{k} \overset{(\Gamma)}{R^{\sigma}_{.\lambda\mu\nu}}$. Using Eq.(A1), the following relation for torsion tensor can also be verified:

$$T^{i}_{.\mu\nu}(x) = h^{i}_{.\lambda}T^{\lambda}_{.\mu\nu} = \frac{1}{2}(h^{i}_{.\mu,\nu} - h^{i}_{.\nu,\mu}) + \frac{1}{2}(\Gamma^{i}_{.j\nu}h^{j}_{.\mu} - \Gamma^{i}_{.j\mu}h^{j}_{.\nu}) \tag{A9}$$